\documentclass[12pt,letterpaper,english]{achemso}
\usepackage[T1]{fontenc}
\usepackage[latin9]{inputenc}
\usepackage{amsmath}
\usepackage{amssymb}
\usepackage{graphicx}
\usepackage{setspace}
\usepackage{esint}
\makeatletter

\title{Attosecond electronic and nuclear quantum photodynamics of ozone
monitored with time and angle resolved photoelectron spectra}
\author{Piero Decleva}
\author{Nicola Quadri}
\affiliation{\label{Dipartimento-di-Scienze}Dipartimento di Scienze Chimiche,
Universita' di Trieste, Via L. Giorgieri 1I - 34127 Trieste, Italy}
\author{Aurelie Perveaux}
\author{David Lauvergnat}
\affiliation{\label{Laboratoire-de-Chimie-1}Laboratoire de Chimie Physique, CNRS,
Université Paris-Sud, F-91405 Orsay, France}
\author{Fabien Gatti}
\author{Benjamin Lasorne}
\affiliation{Institut Charles Gerhardt, CNRS, Université de Montpellier, F-34095
Montpellier, France}
\author{Gábor J. Halász}
\affiliation{\label{Department-of-Information}Department of Information Technology,
University of Debrecen, H-4002 Debrecen, PO Box 400, Hungary}
\author{Ágnes Vibók}
\affiliation{\label{Department-of-Theoretical}Department of Theoretical Physics,
University of Debrecen, H-4002 Debrecen, PO Box 400, Hungary}
\alsoaffiliation{\label{ELI-ALPS,-ELI-HU-Non-Profit}ELI-ALPS, ELI-HU Non-Profit Ltd,
Dugonics tér 13, H-6720 Szeged, Hungary}
\email{vibok@phys.unideb.hu}

\providecommand{\tabularnewline}{\\}

\usepackage{babel}

\usepackage{babel}

\usepackage{babel}

\makeatother

\usepackage{babel}
\begin{document}
\begin{abstract}
Recently we reported a series of numerical simulations proving that
it is possible in principle to create an electronic wave packet and
subsequent electronic motion in a neutral molecule photoexcited by
a UV pump pulse within a few femtoseconds. 

We considered the ozone molecule: for this system the electronic wave
packet leads to a dissociation process. In the present work, we investigate
more specifically the time-resolved photoelectron angular distribution
of the ozone molecule that provides a much more detailed description
of the evolution of the electronic wave packet. We thus show that
this experimental technique should be able to give access to observing
in real time the creation of an electronic wave packet in a neutral
molecule and its impact on a chemical process.
\end{abstract}
\maketitle
\maketitle \maketitle \newpage{}

\section*{Introduction }

Since the advent of femtochemistry remarkable and decisive progress
has been achieved on the experimental front and it is now possible
to monitor electronic motion in the context of attophysics \cite{Dreisler1,Goulielmakis1,Kling,Niikura}.
In other words, electronic wave packets can be created and observed
in real time, which will improve our understanding of fundamental
quantum concepts such as coherence and coherent light-matter interaction
on the time scale of the electrons in a molecule.

Exciting molecules with attosecond XUV light pulses may populate several
electronic states coherently, thus creating an electronic molecular
wave packet. Its evolution will eventually trigger nuclear motion
on a longer timescale via the effective potential created by the electrons
and governing nuclear dynamics. In this context, a crucial challenge
for attosecond sciences is to create specific electronic wave packets
able to induce nuclear motion, e.g. a chemical process, selectively
and efficiently. This should lead, on the long term, to what some
already call attochemistry, where, at each step of a molecular process,
the coupled motions of electrons and nuclei could be controlled on
their natural time scales \cite{kuleff1}. For example, if the attosecond
pulse ionizes the molecule, the hole thus created will move, a process
which is termed charge migration \cite{kuleff1}. This may yield,
in a second step, to selective bond dissociation \cite{kuleff1,Remacle}.
Another possibility is to populate a limited number of electronic
states in the neutral molecule by means of UV subfemtosecond pulses
in order to trigger a selective chemical process. Experimentally,
attosecond pulses are already available in the XUV spectral domain
\cite{Krausz} but few-cycle UV subfemtosecond pulses are expected
to emerge in a near future.

A complete theoretical description of such processes is not a trivial
task: it requires a quantum mechanical description of both the motion
of the electrons and the nuclei in interaction with the external ultrafast
field. In previous studies, we presented a full quantum mechanical
simulation of the excitation of the ozone (neutral) molecule after
excitation by a $3$ fs UV pump pulse \cite{Gabi1,Gabi2,Aurelie1,Aurelie2}.
The central wavelength of the pulse at $260$ nm was selected so as
to create a coherent superposition of only two electronic states:
the ground state, $X$ $({}^{1}A_{1})$, and the excited $B$ ($^{1}B_{2}$)
state \cite{Gabi2}. The ozone molecule was chosen since, for obvious
environmental reasons, its electronic excited states are well-known
and understood \cite{Gabriel1,Schinke1,Schinke2,Schinke3}. In addition
the $B$ state is rather well isolated and, more importantly, the
transition dipole between the $X$ and $B$ state is very large, leading
to the so-called Hartley band in the UV domain that is responsible
for the properties of the ozone layer. As a consequence, exciting
the molecule to the $B$ state does not require very high intensity
(we used a value of $10^{13}$ W/cm$^{2}$), and we can assume that
only this state is populated by the laser pulse. However, it is worth
noting that obtaining such intensities for very short UV pulses remains
an experimental challenge at the moment.

In Ref. \cite{Gabi2}, we investigated the creation of an electronic
wave packet (see Fig. 6 in Ref. \cite{Gabi2}) leading to an oscillation
of the electronic charge density from one O-O bond to the other on
the subfemtosecond time scale (with a period of $0.8$ fs). This wave
packet was thus an alternating superposition of two resonant forms
that are precursors of the two dissociation channels O + O$_{2}$
and O$_{2}$ + O. Upon propagating nuclear wave packets with the Heidelberg
Multi-Configuration Time-Dependent Hartree (MCTDH) package \cite{Heidi1,Heidi2,Heidi3,Heidi4,Heidi5,Heidi6},
we showed that, at the end of the laser pulse, the molecule started
to vibrate (see Fig. 4 in Ref. \cite{Gabi2}). The quantum coherence
between the two electronic states could thus be expected to be destroyed
rapidly due to vibrations, even more so because of the dissociation
outcome making this process irreversible. However, we observed a revival
of coherence after the external field was off, with a time delay corresponding
to a single vibrational period in the $B$ state. This was attributed
to a portion of the wave packet being trapped in the $B$ state around
a shallow potential energy well. Obviously, electronic coherence would
have been preserved longer if the potential energy well of the $B$
state had been deeper. In any case, this revival of quantum coherence
is the signature that the coherent superposition of the two electronic
states is not destroyed as soon as the nuclear motions starts. To
conclude, we showed that it was possible to first create an electronic
wave packet in the bound molecule, which would lead, in a second step,
to the dissociation of the molecule and monitor the whole process
with time-resolved spectroscopy. In principle, one could also expect
to control this process upon manipulating the initial electronic wave
packet via modulating the pump pulse.

From the experimental point of view, a wave packet cannot be observed
as such, or at least not ``directly'' but rather from its consequences
on the photodynamics of the system, via time-resolved observables
obtained from pump-probe spectroscopy techniques. Attosecond XUV probe
pulses can be used to ionize the molecule during the whole process
with a time resolution compatible with the electronic motion\cite{Wollenhaupt,spectra1,spectra2,Krylov1,spanner1,spanner2}.
The resulting time-resolved spectra from both electronic states, $X$
and $B$, will provide precious information about the detailed dynamics
of the system. Our probe pulse is centered around $95$ eV. This high
value generates electrons that are ejected with high velocities. A
sudden approximation can thus be invoked to describe one-photon XUV
ionization \cite{pickup}. In addition, it is desirable that the ionization
process is as instantaneous as possible so that it does not perturb
the electronic motion induced by the pump pulse. In Ref. \cite{Aurelie1},
we calculated the relative ionization probabilities based on an approach
exploiting Dyson orbitals (see Ref. \cite{Aurelie1} for the calculation
of these). Within the sudden approximation regime one can estimate
relative cross sections as the square norms of the Dyson orbitals.
Then, after convolution of the stick photoelectron spectra from $X$
and $B$, we could calculate the time-resolved photoelectron spectrum
(TRPES) as a function of time and photoelectron kinetic energy. This
spectrum clearly exhibited depletion of $X$ and production of $B$
\cite{Aurelie2}.

Now, in order to analyze the wave packet created by the pump pulse
in more detail, it is useful to consider a more accurate and complete
description of the time-resolved photoelectron spectrum, including
both realistic cross sections and angular distributions, and their
photon energy dependence. For instance, molecular frame photoelectron
angular distributions (MFPAD) give access to the shape of the electronic
wave packet \cite{spectra2}. Even photoionization from molecules
that are randomly distributed in terms of their orientation in space
show important dependence on the angle between the polarization axis
of the pump pulse and the direction of the ejected electron. The aim
of the present work is precisely to provide such a time-resolved photoelectron
angular distribution for the dissociation of ozone with the aforementioned
pump pulse. This completes an ab-initio theoretical framework for
the accurate description of pump-probe experiments in small molecules,
represented here by O$_{3}$, able to deal with electronic and nuclear
motion on equal footing, describing the combined electron-nuclear
wave packet.

The outline of the paper is as follows: in the next section we describe
briefly the methods used for quantum chemistry calculations and quantum
dynamics simulations. In the third section, the resulting photoelectron
spectra are presented and discussed. Finally, conclusions provide
an outlook for the future of molecular attophysics.

\section*{Theoretical background}

A molecule such as ozone can be viewed as a collection of $N$ nuclei
and $n$ electrons. Let $\vec{R}$ = $(\vec{R}_{1},\ldots,\vec{R}_{N})$
and $\vec{r}$ = $(\vec{r}_{1},\ldots,\vec{r}_{n})$ denote the position
vectors of the nuclei and the electrons, respectively. Using a semi-classical
approach with respect to the external electromagnetic field and the
so-called dipole approximation, the non-relativistic Coulomb molecular
Hamiltonian operator for the system interacting with a time-dependent
external electric field, $\vec{E}(t)$, reads

\begin{equation}
{H}(\vec{r},\vec{R},t)={T}^{nu}(\vec{R})+{H}^{el}(\vec{r};\vec{R})-\vec{\mu}(\vec{r},\vec{R})\cdot\vec{E}(t)\,,\label{eq:hamiltonian}
\end{equation}
where ${T}^{nu}(\vec{R})$ is the kinetic energy operator of the nuclei,
${H}^{el}(\vec{r};\vec{R})$ the electronic Hamiltonian operator (the
sum of the latter two terms being the field-free molecular Hamiltonian),
and $\vec{\mu}(\vec{r},\vec{R})$ the electric dipole moment of the
molecule.

The time-dependent Schrödinger equation reads 
\begin{eqnarray}
{H}(\vec{r},\vec{R},t)\,\Psi(\vec{r},\vec{R},t) & = & i\hbar\frac{\partial\Psi(\vec{r},\vec{R},t)}{\partial t}\,,\label{eq:tdsfull}
\end{eqnarray}
with $\Psi(\vec{r},\vec{R},t)$ the wave packet of the molecule.

The adiabatic electronic basis functions, $\Phi_{i}(\vec{r};\vec{R})$,
satisfy for each $\vec{R}$ 
\begin{eqnarray}
{H}^{el}(\vec{r};\vec{R})\,\Phi_{i}(\vec{r};\vec{R}) & = & E_{i}^{el}(\vec{R})\,\Phi_{i}(\vec{r};\vec{R})\,,\label{eq:pesadia}
\end{eqnarray}
where $\vec{R}$ are to be viewed as parameters and $E_{i}^{el}(\vec{R})$
play the role of potential energy surfaces for the nuclei.

Here, we consider only a pair of adiabatic electronic states for ozone:
$X$($^{1}A_{1}$), the ground state, and $B$($^{1}B_{2}$), the
Hartley excited state. The total wave function of the molecule can
be expanded as 
\begin{eqnarray}
\Psi(\vec{r},\vec{R},t) & = & \sum_{i=X,B}\Psi_{i}(\vec{R},t)\,\Phi_{i}(\vec{r};\vec{R})\,.\label{eq.expan}
\end{eqnarray}
In the following, we assume the Born-Oppenheimer approximation to
be valid and thus neglect the non-adiabatic couplings between the
two electronic states stemming from the nuclear kinetic energy operator.
The only coupling between $X$ and $B$ is induced by the external
field through the term $-\vec{\mu}_{XB}(\vec{R})\cdot\vec{E}(t)$,
where the transition dipole is defined as $\vec{\mu}_{XB}(\vec{R})=\int\Phi_{B}^{\star}(\vec{r};\vec{R})\vec{\mu}(\vec{r},\vec{R})\Phi_{X}(\vec{r};\vec{R})d\vec{r}$.
We also neglect the diagonal terms involving $\vec{\mu}_{XX}(\vec{R})$
and $\vec{\mu}_{BB}(\vec{R})$ since $\vec{E}(t)$ is an external
field resonant between $X$ and $B$ with respect to the central wavelength
of the spectrum of the pulse.

Thus, the evolution of $\Psi_{X}(\vec{R},t)$ and $\Psi_{B}(\vec{R},t)$
is governed by a set of two coupled equations involving only $E_{X}^{el}(\vec{R})$,
$E_{B}^{el}(\vec{R})$, $-\vec{\mu}_{XB}(\vec{R})\cdot\vec{E}(t)$,
and ${T}^{nu}(\vec{R})$. To solve this set of equations, i.e. to
solve the Schrödinger equation for the nuclei, we use the MCTDH method
\cite{Heidi1,Heidi2,Heidi3,Heidi4,Heidi5,Heidi6,Heidi7}. The nuclear
wave functions are expanded in a basis set of time--dependent functions,
the so--called \emph{single--particle functions} (SPFs), 
\begin{eqnarray}
\Psi(Q_{1},\cdots,Q_{f},t) & = & \sum_{j_{1}}^{n_{1}}\cdots\sum_{j_{f}}^{n_{f}}A_{j_{1},\cdots,j_{f}}(t)\,\prod_{\kappa=1}^{f}\varphi_{j_{\kappa}}^{(\kappa)}(Q_{\kappa},t)\;,\label{eq:expan1}
\end{eqnarray}
where $f$ denotes the number of nuclear degrees of freedom ($Q_{\kappa}$
are single coordinates or groups of coordinates involved in $\vec{R}$).
There are $n_{\kappa}$ SPFs for the $\kappa$th nuclear degree of
freedom. The equations of motion \cite{Heidi1,Heidi2,Heidi3,Heidi4,Heidi5,Heidi6}
for the $A$-coefficients and the SPFs are derived from a variational
principle that ensures optimal convergence.

In this work, $Q_{1},\cdots,Q_{3}$ are (polyspherical) valence coordinates
($R_{1}$ and $R_{2}$, the two bond lengths, and $\alpha$, the angle
between the two bonds). The corresponding expression of the kinetic
energy operator, ${T}^{nu}(R_{1},R_{2},\alpha)$, with zero total
angular momentum can be found in Ref. \cite{Fabien5}. The potential
energy surfaces, $E_{X}^{el}(R_{1},R_{2},\alpha)$ and $E_{B}^{el}(R_{1},R_{2},\alpha)$,
and the transition dipole surface, $\vec{\mu}_{XB}(R_{1},R_{2},\alpha)$,
are those from Schinke and coworkers \cite{Schinke1,Schinke2,Schinke3}.
They are implemented in MCTDH and have already been tested on accurate
applications in spectroscopy \cite{Fabien1,Fabien2,Fabien3,Fabien4}.

The parameters defining $\vec{E}(t)$, the laser pump pulse (see Fig.
1) are: central wavelength at $260$ nm, intensity of $10^{13}$ W/cm$^{2}$,
Gaussian envelope with a full duration at half maximum (FDHM) equal
to $3$ fs. Note that, due to the $C_{2v}$ symmetry of the ozone
molecule at the Franck-Condon (FC) point ($R_{1}=R_{2}=1.275\,\textrm{\AA}$;
$\alpha=116.9^{\circ}$), the $y$-component ($B_{2}$) of the transition
dipole between $X$ and $B$ is the only non-vanisihing one at the
FC point and is thus primarily responsible for the light-induced electronic
transitions. Consequently, the effective polarization axis of the
electric field is $y$.

Further details regarding our calculations -- the (time-independent)
primitive basis sets, the parameters for the complex absorbing potentials,
the refitting of the potential energy and transition dipole surfaces
in a form adapted to MCTDH, and the number of SPFs -- can be found
in previous work, for instance in Sec. 3 of Ref. \cite{Fabien1}.

Starting from the vibrational ground state in the electronic ground
state $X$, MCTDH calculations will generate $\Psi_{X}(\vec{R},t)$
and $\Psi_{B}(\vec{R},t)$ at any subsequent time. Assuming that only
the $B$ electronic state is populated by the laser pulse (see Fig.
1), the total molecular wave packet (see Eq. \ref{eq.expan}) can
be constructed provided the corresponding adiabatic electronic wave
functions are known.

Thus, with this approach, we can obtain in principle the full electronic
and vibrational wave packet (note again that we only consider the
case where the total angular momentum is equal to $0$). However,
this quantity cannot be observed directly in actual experiments and
we need a time-resolved property that will characterize the time evolution
of the system: the TRPES for instance, which can be measured and compared
to calculations. The procedure that we used to compute this quantity
is explained below.

As a first approximation, we can consider that the early stages of
the process will be dominated by the behavior of the wave packet at
the FC point, $\vec{R}_{FC}$. The corresponding renormalized density
matrix of the molecule at the FC point (see Sec. II B of Ref. \cite{Gabi2}
for further details) reads, for $i,\, i^{'}=X,\, B$, 
\begin{eqnarray}
\rho_{ii^{'}}(t) & = & \frac{\Psi_{i}^{\star}(\vec{R}_{FC},t)\Psi_{i\text{'}}(\vec{R}_{FC},t)}{\sum_{l=X,B}\left|\Psi_{(l)}(\vec{R}_{FC},t)\right|^{2}}\label{eq:totdensity}
\end{eqnarray}
Note that such local populations of $X$ and $B$ are not classical
quantities but extracted from the actual quantum wave packets.

Assuming a ``stationary'' picture, the approximate photoelectron
spectra from either $X$\cite{Ohtsuka} or $B$ at the FC point appear
as stick spectra, 
\begin{equation}
I_{k}(\epsilon)=\sum_{i}I_{ik}\delta(\epsilon-\epsilon_{ik})\,,\label{eq:stick}
\end{equation}
where $\epsilon$ is the kinetic energy (KE) of the ejected electron,
$i=X$ or $B$, and $k$ is used to label the various cation states.
$\epsilon_{ik}$ are the corresponding peaks appearing in the spectra.
They satisfy 
\begin{equation}
\epsilon_{ik}=E_{photon}-IP_{ik}\;\;\;\;\; IP_{ik}=E_{k}-E_{i}\,,\label{eq:epsilon}
\end{equation}
where $E_{photon}$ denotes the energy of the probe photon, 95 eV
here. $E_{i}$ are the energies of the $X$ and $B$ states at the
FC geometry, $E_{k}$ the energies of the cation that can be populated
by the photon at the same geometry, and $IP_{ik}$ are the relative
ionization potentials. Our calculations show that $19$ cation states
can be populated (up to about $20$ eV above the $X$ state) \cite{Aurelie2}.
For the calculation of the peak intensities, $I_{ik}$, we adopt an
approach based on Dyson orbitals \cite{Aurelie1}. The latter are
defined as 
\begin{eqnarray}
\phi_{i,k}^{Dyson}(\vec{r};\vec{R}) & = & \sqrt{n}\int d\vec{r}_{2}\ldots d\vec{r}_{n}\Phi_{i}^{el}(\vec{r}=\vec{r_{1}},\vec{r}_{2},\ldots,\vec{r}_{n};\vec{R})\nonumber \\
 & \times & \Phi_{k}^{cat\star}(\vec{r}_{2},\ldots,\vec{r}_{n};\vec{R})\,,\label{eq:dyson}
\end{eqnarray}
where $\Phi_{i}^{el}$ are the electronic functions of the neutral
molecule as defined above and $\Phi_{k}^{cat}$ the electronic functions
of the cation. We calculated Dyson norms at the FC point (see Ref.
\cite{Gabi2}) at the CASSCF(17,12)/aug-cc-pVQZ (no state average)
level of theory for the cation wave functions and CASSCF(18,12)/aug-cc-pVQZ
(no state average) for the neutral wave functions with the MOLPRO
quantum chemistry package \cite{Molpro}. The energies of the neutral
and the cation were further refined with MRCI-SD(Q) calculations,
including Davidson corrections, and based on the previous CASSCF references.

If a sudden approximation is assumed, the squares of the Dyson norms,
$\langle\phi_{i,k}^{Dyson}\vert\phi_{i,k}^{Dyson}\rangle$, are proportional
to the relative ionization probabilities $I_{ik}$. Ionization potentials
and $I_{ik}=\langle\phi_{i,k}^{Dyson}\vert\phi_{i,k}^{Dyson}\rangle$
are reported in \ref{tab:Ab-initio-ionization}. The corresponding
stick spectrum is displayed in Figure 2. To obtain the energy resolved
spectra we convoluted the stick spectra with a Gaussian envelope function
$G(\varepsilon)$ to mimic the bandwidth of the XUV probe pulse, 
\begin{equation}
I_{k}(\varepsilon)=\sum_{j}G_{jk}(\varepsilon)I_{jk}\;\;\;\;\;\; G_{jk}(\varepsilon)=\frac{1}{\sigma\sqrt{2\pi}}e^{-\frac{(\varepsilon-\varepsilon_{jk})^{2}}{2\sigma^{2}}}.\label{eq:dyn6}
\end{equation}
Here $\sigma$ is the standard deviation of the intensity: $\sigma=1.5$
eV for a probe pulse of FDHM equal to $500$ as.

Let us now consider the full photoionization dynamics. Assuming a
randomly oriented molecular sample, the differential cross section
in the laboratory frame (LF) coordinate system is given by the following
expression:

\begin{equation}
\frac{d\sigma_{jk}(\varepsilon_{jk})}{d\Omega}=\frac{\sigma_{jk}(\varepsilon_{jk})}{4\pi}[1+\beta_{jk}(\varepsilon_{jk})P_{2}(\cos\theta)]\label{eq:cross}
\end{equation}
where $P_{2}(\cos\theta)=\frac{1}{2}(3\cos^{2}\theta-1)$ is the second
order Legendre polynomials and $\theta$ is the angle between the
direction of the electron momentum and the polarization of the electric
field. $\Omega$ is the angle relative to electron emission momentum
in the LF system and the two energy dependent parameters are $\sigma_{jk}$
(partial cross section) and $\beta_{jk}$ (asymmetry parameter). (The
LF system defines the experiment i.e. the direction of the polarization
and propagation of light as well as the direction of electron detection.
The reference system is the molecular frame (MF) system in which the
molecule is fixed and the electronic structure, transition dipole
moment etc. calculations are performed.)

%  revised text 30.05.2016  Piero

Calculation of $\sigma$ and $\beta$ parameters require an explicit
description of the continuum wave function for the final state. Neglecting
interchannel coupling effects, generally very small far from thresholds,
a single channel approximation of the form 
\begin{equation}
\Psi_{k,\vec{\kappa}}^{(-)}=\textsl{A}\Phi_{k}^{cat}\varphi_{\vec{\kappa}}^{(-)}\label{SC}
\end{equation}
is generally quite accurate. Here $\varphi_{\vec{\kappa}}^{(-)}$
describes an electron with asymptotic momentum $\vec{\kappa}$ (and
incoming wave boundary conditions, appropriate for photoionization),
and $\textsl{A}$ describes antisymmetrization and proper symmetry
couplings. Actually it is computationally easier to work in an angular
momentum basis, employing eigenstates 
\begin{equation}
\Psi_{k,\varepsilon lm}=\textsl{A}\Phi_{k}^{cat}(\vec{r}_{1},\ldots,\vec{r}_{N-1})\varphi_{\varepsilon lm}(\vec{r}_{N})\label{SC1}
\end{equation}
where the continuum wavefunctions $\varphi_{\varepsilon lm}$ are
characterized by suitable asymptotic conditions, in our case K-matrix
boundary conditions, defined as 
\begin{equation}
\varphi_{\varepsilon lm}(\vec{r})\rightarrow\sum_{l'm'}(f_{l'}(\kappa r)\delta_{l'l}\delta_{m'm}+g_{l'}(\kappa r)K_{l'm',lm})Y_{l'm'}\label{SC2}
\end{equation}
which has the advantage of working with real wave functions. Here
$f_{l}$ and $g_{l}$ are regular and irregular coulomb functions.
The $\varphi_{\varepsilon lm}$ so obtained can be transformed to
incoming wave boundary conditions and then to linear asymptotic momentum
by standard transformation \cite{Chandra} 
\begin{equation}
\varphi_{\varepsilon lm}^{(-)}=\sum_{l'm'}\varphi_{\varepsilon l'm'}(1+iK)_{l'm',lm}^{-1}\label{SC3}
\end{equation}
\begin{equation}
\varphi_{k,\vec{\kappa}}^{(-)}=\frac{1}{\sqrt{\mu}}\sum_{l'm'}i^{l}e^{-i\sigma_{l}}Y_{lm}(\hat{\kappa})\varphi_{\varepsilon lm}^{(-)}\label{SC4}
\end{equation}
The same transformation can be directly applied to the transition
dipole moments. The many-particle transition dipole moment 
\begin{equation}
D_{ik;lm\gamma}(\varepsilon)=\langle\textsl{A}\Phi_{k}^{cat}\varphi_{\varepsilon lm}|D_{\gamma}|\Phi_{i}^{el}\rangle\label{D_1}
\end{equation}
reduces to the single particle moment involving the Dyson orbital
(\ref{eq:dyson}) 
\begin{equation}
D_{ik;lm\gamma}(\varepsilon)=\langle\varphi_{\varepsilon lm}|d_{\gamma}|\phi_{i,k}^{Dyson}\rangle\label{D_2}
\end{equation}
plus an additional term (conjugate term) which is generally small
and is usually neglected \cite{Aurora1}. Here $\gamma$ is the Cartesian
component of the dipole, $D$ and $d$ are the many-particle and the
single particle dipole operators.

From dipole moments (and the $K$-matrix) $\sigma_{jk}(\varepsilon)$
and $\beta_{jk}(\varepsilon)$, as well as any angular distribution
from oriented molecules, can be computed according to well known formulas
\cite{Chandra}.

In our formulation, the continuum wave function (\ref{SC1}) is computed
as an eigenfunction of the Kohn-Sham Hamiltonian defined by the initial
state electron density $\rho$ 
\begin{equation}
h_{KS}\varphi_{\varepsilon lm}=\varepsilon\varphi_{\varepsilon lm}\label{HK}
\end{equation}
\begin{equation}
h_{KS}=-\frac{1}{2}\Delta+V_{eN}+V_{C}(\rho)+V_{XC}(\rho)\label{HK1}
\end{equation}
where $V_{eN}$ is the nuclear attraction potential, $V_{C}$ the
coulomb potential and $V_{XC}$ the exchange-correlation potential
defined in terms of the ground state density $\rho$. The latter is
obtained from a conventional LCAO SCF calculation, employing the ADF
program with a DZP basis. A special basis is employed for the continuum
solutions of (\ref{HK}). Primitive basis functions are products of
a $B$-spline radial function \cite{Bachau,deboor} times a real spherical
harmonic 
\begin{equation}
\chi_{ilm}(r,\theta,\phi)=\frac{1}{r}B_{i}(r)Y_{lm}(\theta,\phi)\label{basis_1}
\end{equation}
The full basis comprises a large one-center expansion on a common
origin, with long range $R_{max0}$, and large maximum angular momentum,
$L_{max0}$. This is supplemented by additional functions centered
on the nuclei, of very short range, $R_{maxp}$, and small angular
momenta $L_{maxp}$. A short range is necessary to avoid almost linear
dependence of the basis, which spoils the numerical stability of the
approach. Despite the very limited number of LCAO functions these
choices ensure a very fast convergence of the calculated quantities.
The basis is then fully symmetry adapted.

The calculation of continuum eigenvectors is performed at any selected
electron kinetic energy by the Galerkin approach originally proposed
in ref. \cite{fischer1989} and the generalized to the multichannel
case \cite{brosolo1992cp,brosolo1992jpb}. From the energy independent
Hamiltonian $H$ and overlap $S$ matrices continuum vectors are obtained
as eigenvectors of the energy dependent matrix $A(E)=H-ES$ with eigenvalues
very close to zero. These give the correct number of independent open
channel solutions, and are efficiently obtained by block inverse iteration,
since they are separated by large gaps from the rest of the spectrum.
Actually the more stable form $A^{+}A$ is currently employed \cite{brosolo1992cpc}.
Final normalization to $K$-matrix boundary conditions is obtained
by fitting the solutions to the analytical asymptotic form at the
outer boundary $R_{max0}$.

In the present calculation the LB94 $V_{XC}$ potential \cite{Leeuwen1994}
was employed, due to the correct asymptotic behavior, important in
photoionization. Parameters were $L_{max0}=12$, $R_{max0}=25.0$
a.u., with 135 B-splines of order 10, $L_{maxp}=2$, $R_{maxp}=1.50$
a.u. for the O atoms, for a total of 23013 basis functions.

Such an approach, called static-DFT proves in general remarkably accurate
for the description of cross sections and angular distributions \cite{Bachau,Decleva1,Stener1}.
In conjunction with the Dyson orbital formulation it is able to describe
ionization involving multiconfigurational initial and final cationic
states \cite{Aurora1,Aurora2}. We refer to previous work for details
of the implementation \cite{Bachau,Toffoli1}. $\sigma_{jk}$ and
$\beta_{jk}$ are obtained on a dense electron KE $\varepsilon_{jk}$
grid, so that the value at any KE dictated by the given photon energy
can be accurately obtained by interpolation. With these the angularly
resolved photoelectron intensity becomes:

\begin{equation}
I_{k}(\varepsilon,\theta)=\sum_{j}G_{jk}(\varepsilon_{jk})\frac{\sigma_{jk}(\varepsilon_{jk})}{4\pi}[1+\beta_{jk}(\varepsilon_{jk})P_{2}(\cos\theta)].\label{eq:ang-int}
\end{equation}
Applying the same convolution procedure as in Eq. 9 of Ref. \cite{Aurelie2}
we arrive to the appropriate formula of the angle resolved photoelectron
spectrum:

\begin{equation}
I(\varepsilon,\theta,\tau)=\sum_{k}\rho_{kk}(\tau)I_{k}(\varepsilon,\theta).\label{eq:final}
\end{equation}
Here the $\rho_{kk}(\tau)$ comes from eq. \ref{eq:totdensity} and
from now on the above expression (eq. \ref{eq:final}) will serve
as our working formula in the forthcoming part of the paper.

\begin{figure}
\begin{centering}
\includegraphics[width=0.5\textwidth]{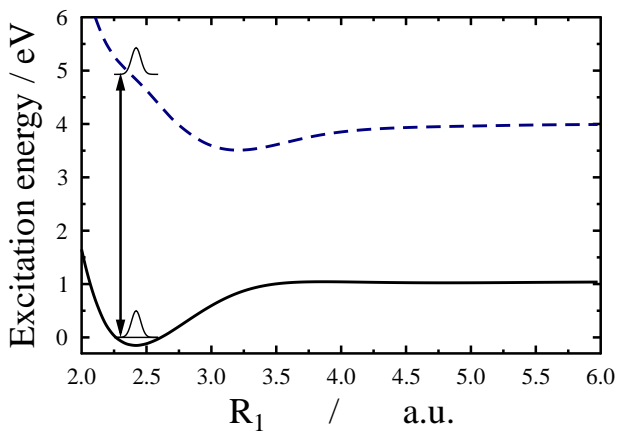}
\par\end{centering}

\caption{Potential energy cut of the ozone molecule as a function of the dissociation
coordinate, $R_{1}$: ground state ($X$, solid line) and Hartley
state ($B$, dashed line), the arrow denotes the excitation of the
$B$ state. The other bond is fixed at $R_{2}=2.43$ a.u. and the
bond angle $\alpha$$=117^{\circ}$. }
\end{figure}

\begin{figure}
\begin{centering}
\includegraphics[width=0.5\textwidth]{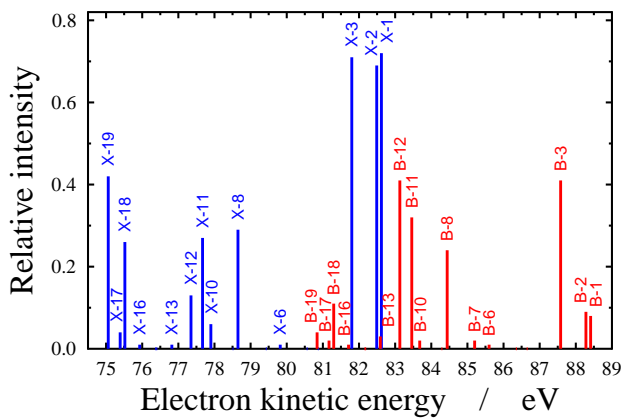}
\par\end{centering}

\caption{Stick photoelectron spectra from $X$ (blue) or $B$ (red) as functions
of the energy of the ejected electron for a probe photon at $95$
eV. Cation states (see Table 1) are labeled according to the order
given in Ref. \cite{Ohtsuka}; our calculations give $E_{15}<E_{14}$
and $E_{18}<E_{17}$, which is why $B-18$ is before $B-17$.}
\end{figure}

\begin{figure}
\begin{centering}
\includegraphics[width=1\textwidth]{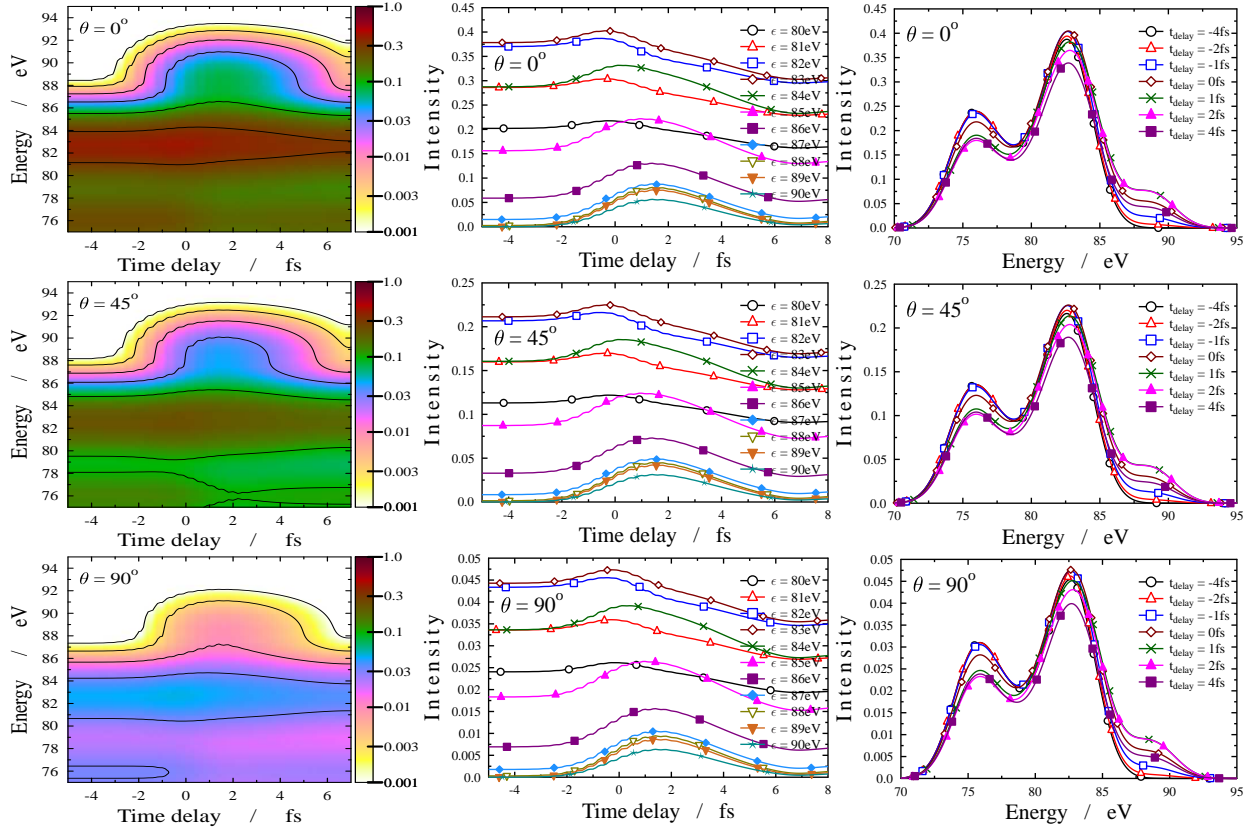}
\par\end{centering}

\caption{Angle resolved photoelectron spectrum (ARPES). First column: ARPES
(logarithmic scale) as a function of the time delay (horizontal axis)
and energy of the ejected electrons (vertical axis). The different
panels correspond to different $\theta$ orientation angle ($\theta$
is the angle between the direction of the electron momentum and the
polarization of the electric field). The intensity of the ejected
electrons are coded by colors according to the scale on the right
side. Second column: One dimensional cuts for the intensity of the
ejected electrons via time delay with fixed $\theta$ and $\epsilon$.
Third column: One dimensional cuts for the intensity of the ejected
electrons via energy with fixed $\theta$ and $t_{delay}$.}
\end{figure}

\begin{figure}
\begin{centering}
\includegraphics[width=1\textwidth]{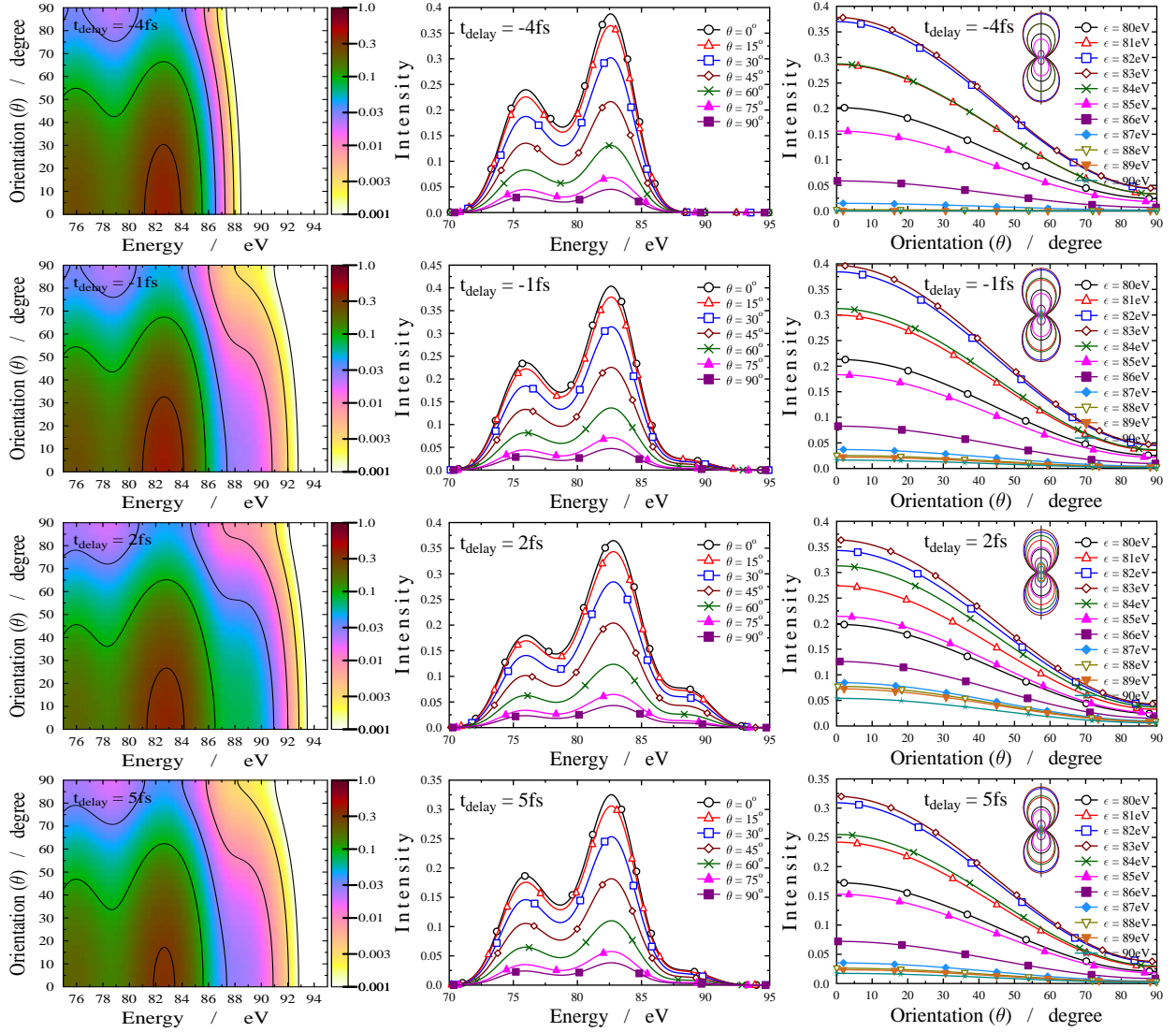}
\par\end{centering}

\caption{Angle resolved photoelectron spectrum (ARPES). First column: ARPES
(logarithmic scale) as a function of the energy of the ejected electrons
(horizontal axis) and orientation angle $\theta$ ($\theta$ is the
angle between the direction of the electron momentum and the polarization
of the electric field) (vertical axis). The different panels correspond
to different time delays between the pump and probe pulses. The intensity
of the ejected electrons are coded by colors according to the scale
on the right side. Second column: One dimensional cuts for the intensity
of the ejected electrons via energy with fixed $t_{delay}$ and $\theta$.
Third column: One dimensional cuts for the intensity of the ejected
electrons via electron emission orientation with fixed $t_{delay}$
and $\epsilon$.}
\end{figure}

\begin{figure}
\begin{centering}
\includegraphics[width=1\textwidth]{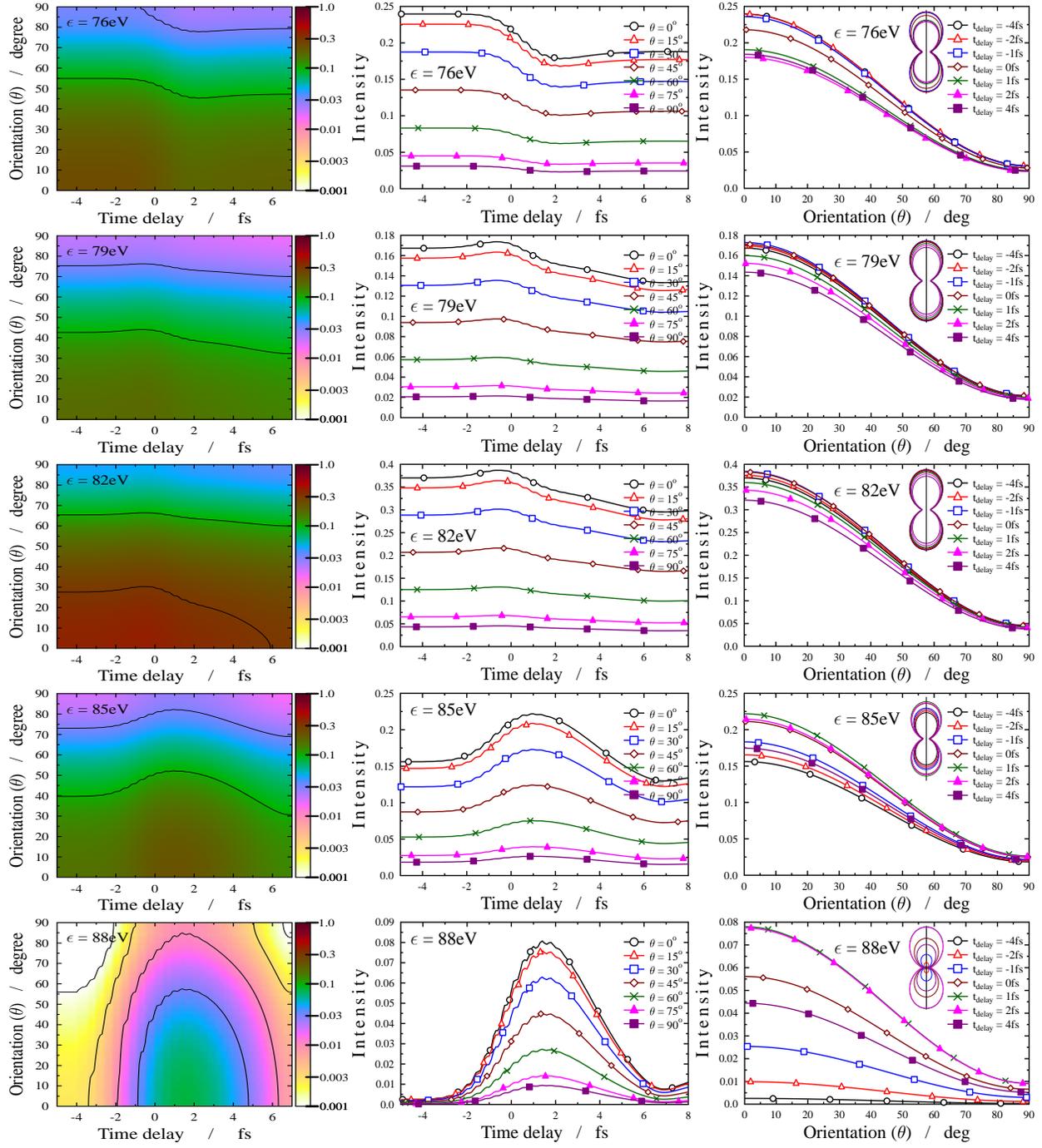}
\par\end{centering}

\caption{Angle resolved photoelectron spectrum (ARPES). First column: ARPES
(logarithmic scale) as a function of the time delay $t_{delay}$ (horizontal
axis) and orientation angle $\theta$ ($\theta$ is the angle between
the direction of the electron momentum and the polarization of the
electric field) (vertical axis). The different panels correspond to
different energies of the ejected electrons. The intensity of the
ejected electrons are coded by colors according to the scale on the
right side. Second column: One dimensional cuts for the intensity
of the ejected electrons via time delay with fixed $\epsilon$ and
$\theta$. Third column: One dimensional cuts for the intensity of
the ejected electrons via electron emission orientation with fixed
$E$ and $t_{delay}$. }
\end{figure}

\begin{table}
\begin{centering}
\begin{tabular}{|c|c|c|c|c|c|}
\hline 
 & cation states ($j$)  & $E_{j}-E_{X}/$eV  &  $I_{ik}(X)$  & $E_{j}-E_{B}/$eV  & $I_{ik}(B)$\tabularnewline
\hline 
\hline 
1  & ($1^{2}A_{1}$)  & 12.38  & 0.72  & 6.59  & 0.08\tabularnewline
\hline 
2  & ($1^{2}B_{2}$)  & 12.51  & 0.69  & 6.72  & 0.09\tabularnewline
\hline 
3  & ($1^{2}A_{2}$)  & 13.20  & 0.71  & 7.42  & 0.41\tabularnewline
\hline 
4  & ($1^{2}B_{1}$)  & 14.14  & 0.00  & 8.36  & 0.00\tabularnewline
\hline 
5  & ($2^{2}A_{2}$)  & 14.45  & 0.00  & 8.66  & 0.00\tabularnewline
\hline 
6  & ($2^{2}B_{2}$)  & 15.18  & 0.01  & 9.40  & 0.01\tabularnewline
\hline 
7  & ($2^{2}A_{1}$)  & 15.58  & 0.00  & 9.80  & 0.02\tabularnewline
\hline 
8  & ($2^{2}B_{1}$)  & 16.35  & 0.29  & 10.56  & 0.24\tabularnewline
\hline 
9  & ($3^{2}A_{2}$)  & 16.50  & 0.00  & 10.72  & 0.00\tabularnewline
\hline 
10  & ($3^{2}B_{1}$)  & 17.10  & 0.06  & 11.32  & 0.02\tabularnewline
\hline 
11  & ($3^{2}A_{1}$)  & 17.33  & 0.27  & 11.54  & 0.32\tabularnewline
\hline 
12  & ($3^{2}B_{2}$)  & 17.65  & 0.13  & 11.87  & 0.41\tabularnewline
\hline 
13  & ($4^{2}B_{2}$)  & 18.18  & 0.01  & 12.41  & 0.03\tabularnewline
\hline 
14  & ($4^{2}A_{2}$)  & 18.64  & 0.00  & 12.85  & 0.00\tabularnewline
\hline 
15  & ($4^{2}B_{1}$)  & 18.61  & 0.00  & 12.83  & 0.00\tabularnewline
\hline 
16  & ($4^{2}A_{1}$)  & 19.07  & 0.01  & 13.29  & 0.01\tabularnewline
\hline 
17  & ($5^{2}B_{2}$)  & 19.61  & 0.04  & 13.83  & 0.02\tabularnewline
\hline 
18  & ($5^{2}A_{1}$)  & 19.48  & 0.26  & 13.70  & 0.11\tabularnewline
\hline 
19  & ($6^{2}B_{2}$)  & 19.94  & 0.42  & 14.16  & 0.04\tabularnewline
\hline 
\end{tabular}
\par\end{centering}

\caption{\emph{\label{tab:Ab-initio-ionization}Ab initio }ionization potentials
(MRCI-SD(Q) level of theory) and $I_{ik}$, the squares of the Dyson
norms (CASSCF/aug-cc-pVQZ level of theory) with respect to either
$X$ or $B$ at the FC point. The energy difference between the $X$
and $B$ states is $5.78$ eV. (Experimental ionization potentials
and further theoretical values can be found for comparison in Ref.
\cite{Ohtsuka}.)}
\end{table}

\section*{Results and Discussion}

Figure 3 displays the intensity of the ejected electrons as a function
of energy and time delay between the pump and probe pulses for three
different fixed values of the orientation angle, $\theta$. It can
be seen that the ionization probability is larger for smaller angles.
For $\theta>45^{\circ}$ it is drastically reduced. At early times,
when $t_{delay}<-2$ fs, ionization can only take place from the ground
state, $X$. Here, two clearly distinct high intensity bands are observed
within the $75\,-\,78$ eV and the $80\,-\,85$ eV energy intervals.
These are consistent with the large Dyson norms calculated between
the $X$ state of the neutral and some of the states of the cation
(see Table 1). In particular, large Dyson norms are found between
$X$ and the 1st ($0.72$), 2nd ($0.69$), 3rd ($0.71$), 8th ($0.29$),
11th ($0.27$), 18th ($0.26$), and 19th ($0.42$) cationic states.
The corresponding ionization potential values for these lie within
($12.38\,-\,13.2$) eV and ($16.35\,-\,19.94$) eV, thus resulting
in two well separated energy regions, $\sim(80\,-\,85)$ eV and $\sim(75\,-$
$78)$ eV. However, from $t_{delay}=-2$ fs on, the pattern becomes
richer due to ionization appearing from $B$ as well. The explicit
consequence of this is a new band that appears around $88$ eV in
the $t_{delay}=0\,-4$ fs time interval. This indicates that the $B$
state starts to be populated, owing to the large value of the Dyson
norm between $B$ and the 3rd cationic state ($0.41$). In addition,
significant ionization is achieved from $B$ to the 8th ($0.24$),
11th ($0.32$), and 12th ($0.41$) cationic states, which corresponds
to the energy band around ($80\,-\,85$) eV in the $t_{delay}=0\,-2$
fs time interval. Simultaneously, for $t_{delay}>0$ fs the $X$ electronic
state slowly depletes, thus providing fewer electrons ejected from
the ground state, which results in smaller intensity values (see the
color in the $75\,-\,78$ eV energy region). The structure of the
figures at larger angles ($\theta>45^{\circ}$) are quite similar
to the former ones, but the colors are much lighter due to lower intensities,
reflecting that large orientation angles are much less likely to be
involved efficiently in the ionization.

The above findings are confirmed on Figure 4 and Figure 5, where the
same results are presented differently. On Figure 4, the electron
emission orientation is given against the energy of the ejected electrons
at several consecutive times. We observe that, up to $t_{delay}=-1$
fs, only two energy regions, $(75\,-\,78)$ eV and $(81\,-\,84)$
eV, exhibit significant intensity. They correspond to ionization taking
place from $X$ only. Ionization occurring from $B$, once $t_{delay}>-2$
fs, is characterized by the third band that appears around $88$ eV
and disappears slowly beyond $t_{delay}>4$ fs. Within the $t_{delay}=1\,-\,2$
fs time interval, the strengthening of the middle band reflects the
combined impact of ionization occurring from both states together.
Again, one clearly sees that, as a general trend, the intensity decreases
monotonically as the angle between the ejected electrons and the direction
of the polarization increases.

In Figure 5, the electron emission orientation is plotted as a function
of the time delay for several fixed electron energy values. Again,
one observes large intensities in the $(75\,-\,77)$ eV energy region
and $t_{delay}<0$ fs time interval for small orientation angles.
The latter corresponds to the lack of population of the $B$ state
resulting in ionization taking place only from the $X$ state. For
$t_{delay}>0$ fs, the decrease of the intensity indicates depletion
of the $X$ state. For $\epsilon>80$ eV, a joint effect of ionizations
from $X$ and $B$ is observed, more substantially from $X$. Again,
the shape and the structure of the band for $\epsilon>85$ eV and
$t_{delay}=(-2)\,-\,6$ fs is typical of ionization occurring from
$B$.

From Figure 5 it also appears that the angular distribution is strongly
peaked along the probe field polarization, which is consistent with
a high $\beta$ value, close to two, for all ionizations. This is
not surprising because of the high photon energy of the probe, $95$
eV, which implies high kinetic energy of the outer valence ionized
electrons, typically characterized by high $\beta$ values, similar
for all ionizations.

Finally the oscillatory patterns appearing in Figures 3 and 5 are
clear fingerprints of the time dependence of the external electric
field. Specifically, the pump pulse is a few-cycle pulse of width
3 fs and period 0.87 fs, centered around 260 nm (4.8 eV) in the deep
UV (UV-C) domain and therefore its oscillation is faster than the
nuclear motion. 

In summary, the most representative signal is perhaps the upper-right
panel in Figure 3 (intensity against electron kinetic energy at different
time delays for $\theta=0^{\circ}$). It is clear that the largest
temporal change in the spectrum is associated with the highest kinetic
energies, from 86 to 89 eV, which are exclusively emitted from the
$B$ state, where the intensity increases significantly just after
the pump pulse. Correspondingly, the decrease of the intensity after
the pump is most evident in the low kinetic energy region, from 75
to 78 eV, due to the depleting of the $X$ state, which is the dominant
contribution in this energy window.

\section*{Conclusions}

A numerical simulation protocol has been developed for describing
the electron dynamics of the ozone molecule in the Franck-Condon region
involving only the ground ($X$) and Hartley ($B$) electronic states
in the dynamics. Assuming isotropic initial distribution for the molecular
ensemble, angle resolved photoelectron spectra have been calculated
for various time delays between the pump that creates the wave packet
(coherent superposition of $X$ and $B$) and the probe that ionizes
from either $X$ or $B$. This physical quantity can be measured in
actual experiments and compared to our calculations. 

The present results are very encouraging and call for further improvements
concerning the accuracy of the dynamics simulations. Therefore, our
future aim is to perform more realistic simulations upon going beyond
the presently assumed limiting hypotheses: isotropic initial distribution
and populations extracted at the FC geometry only. This will be manifested
by two significant changes in the numerical protocol:\emph{ i)} after
the pump pulse is off alignment of the molecular ensemble will be
assumed; \emph{ii) }instead of performing calculations at a single
FC geometry, several other nuclear geometries will be involved in
the FC region where the nuclear density has significant value too.

We stress again that given the dipole matrix elements and K-matrix,
all photoionization observables can be computed, like photoionization
from fixed-in-space molecules (MFPADS) or partially oriented molecules,
as well as suitable averages over final detector energy and angle
resolution \cite{Stener1}, to accurately describe any specific experimental
setup. Actually the 95 eV pulse employed in the present study was
suggested by an experimental colleague. With hindsight angular distribution
from unoriented molecules turn out not to be very informative, given
the $\beta$ values close to $2$ for all final states at this relatively
large photon energy. Working at lower energies would produce larger
anisotropies. Moreover working with oriented molecules, which is a
goal actively pursued in such studies, would further much enhance
anisotropies, different for each initial and final state.

The present numerical simulations clearly indicate that angle and
time resolved photoelectron spectra can be used in molecular attophysics
to characterize the creation of an electronic wave packet in a neutral
molecule on the subfemtosecond time scale. We expect our computational
study to be followed by experiments showing similar results.

As the number of experimental choices is quite large, we found it
important to set up a fully ab-initio general formulation that will
accommodate any specific experimental setup. We look forward to upcoming
experiments to validate the theoretical framework provided here.

\section*{Acknowledgements}

The authors thank H.-D. Meyer for very helpful discussions about the
MCTDH calculations. P.D. and Á.V. acknowledge the supports from the
CORINF and from the COST action CM1204 XLIC.

\section*{Author contributions statement }

P.D., B.L., and A.V. initiated the concept of the calculations. P.D.,
G.J.H. and D.L. conducted the calculations. G.J.H. prepared the figures.
F.G., B.L., P.D. and A.V. wrote the manuscript. All authors analyzed
the results and reviewed the manuscript.

\section*{Additional information }

\textbf{Competing financial interests.} On behalf of all coauthors,
A. V. declares no competing financial interests. 
\end{document}